\documentclass{article}
\usepackage{times}
\begin{document}
\title{Duality and Quantum Mechanics}
\author{Jos\'e M. Isidro\\
Instituto de F\'{\i}sica Corpuscular (CSIC--UVEG)\\
Apartado de Correos 22085, Valencia 46071, Spain\\
{\tt jmisidro@ific.uv.es}}

\maketitle

\begin{abstract}
This {\it Festschrift}\/ in honour of J. A. de Azc\'arraga\footnote{To appear in the proceedings of {\it Symmetries in Gravity and Field Theory}, Workshop in honour of Prof. J.A. de Azc\'arraga, june 9--11, 2003, Salamanca, Spain.} gives an introduction to 
the concept of duality, {\it i.e.}, to the relativity of the notion of a quantum, 
in the context of the quantum mechanics of a finite number of degrees of freedom.
Although the concept of duality arises in string and M--theory, Vafa
has argued that it should also have a counterpart in quantum mechanics, before moving 
on to second quantisation, fields, strings and branes. We illustrate our 
analysis with the case when classical phase space is complex projective 
space,  but our conclusions can be generalised to other complex, symplectic phase spaces, 
both compact and noncompact.

\end{abstract}

\tableofcontents

\section{Introduction}\label{intro}

{}Fibre bundles \cite{LIBAZCA} are powerful tools to formulate the gauge theories 
of fundamental interactions and gravity. The question arises whether or not 
quantum mechanics may also be formulated using fibre bundles.
Important physical motivations call for such a formulation.

In quantum mechanics one aims at constructing a Hilbert--space vector bundle 
over classical phase space. In geometric quantisation 
this goal is achieved in a two--step process that can be very succintly summarised 
as follows. One first constructs a certain holomorphic line bundle (the {\it quantum line
bundle}\/) over classical phase space. Next one identifies certain sections of this line bundle  
as defining the Hilbert space of quantum states. Alternatively one may skip 
the quantum line bundle and consider the one--step process of directly constructing a 
Hilbert--space vector bundle over classical phase space. Associated with this vector bundle 
there is a principal bundle whose fibre is the unitary group of Hilbert space.

Textbooks on quantum mechanics \cite{HOLLAND} usually  deal with the case when this 
Hilbert--space vector bundle is trivial. Such is the case, 
{\it e.g.}, when classical phase space is contractible to a point. However, 
it seems natural to consider the case of a nontrivial bundle as well. 
Beyond a purely mathematical interest, important physical issues that go 
by the generic name of {\it dualities} \cite{VAFA} motivate 
the study of nontrivial bundles. 

Triviality of the Hilbert--space vector bundle implies that the transition 
functions all equal the identity of the structure group. In passing from 
one coordinate chart to another on classical phase space, vectors on the 
fibre are acted on by the identity. Since these vectors are quantum 
states, we can say that all observers on classical phase space are quantised 
in the same way. This is no longer the case on a nontrivial vector bundle, 
where the transition functions are different from the identity.
As opposed to the previous case, different neighbourhoods 
on classical phase space are quantised independently and, possibly, differently. 
The resulting quantisation is only local on classical phase space, instead of global.
This reflects the property of local triviality satisfied by all fibre bundles.

Given a certain base manifold and a certain fibre, the trivial bundle over the given base 
with the given fibre is unique. This may mislead one to conclude that quantisation 
is also unique, or independent of the observer on classical phase space. In 
fact the notion of duality points precisely to the opposite conclusion, {\it i.e.}, 
to the nonuniqueness of the quantisation procedure and to its dependence on the observer 
\cite{VAFA}. 

Clearly a framework is required in order to accommodate dualities within quantum mechanics 
\cite{VAFA}. Nontrivial Hilbert--space vector bundles over classical phase space provide 
one such framework. They allow for the possibility of having different, nonequivalent quantisations,
as opposed to the uniqueness of the trivial bundle. However, although nontriviality 
is a necessary condition, it is by no means sufficient. A flat connection on a nontrivial 
bundle would still allow, by parallel transport, to canonically identify the Hilbert--space 
fibres above different points on classical phase space. This identification would depend 
only on the homotopy class of the curve joining the basepoints, but not on the curve itself. 
Now flat connections are characterised by {\it constant}\/ transition functions 
\cite{LIBAZCA}, 
this constant being always the identity in the case of the trivial bundle. 
Hence, in order to accommodate dualities, we will be looking for {\it nonflat}\/ connections. 
We will see presently what connections we need on these bundles.

This article is devoted to constructing nonflat Hilbert--space vector bundles over classical 
phase space. Our notations are as follows. ${\cal C}$ will denote a complex $n$--dimensional, 
connected, compact classical phase space, endowed with a symplectic form $\omega$ 
and a complex structure ${\cal J}$. We will assume that $\omega$ and ${\cal J}$ 
are compatible, so holomorphic coordinate charts on ${\cal C}$ will also be 
Darboux charts. We will primarily concentrate on the case when ${\cal C}$ is 
complex projective space ${\bf CP}^n$. Its holomorphic tangent bundle will be 
denoted $T({\bf CP}^n)$. The tautological line bundle $\tau^{-1}$ over ${\bf CP}^n$ 
and its dual $\tau$ will also be considered. The Picard group of ${\cal C}$ 
will be denoted ${\rm Pic}\,({\cal C})$. Towards the end of this article we will also 
consider the infinite--dimensional projective space ${\bf CP}({\cal H})$, 
corresponding to complex, separable, infinite--dimensional Hilbert space ${\cal H}$.

{}Finally we would like to draw attention to refs. \cite{MARMO, MACFARLANE, 
SANTANDER, GADELLA, OLMO, MATONE, ME}, 
where issues partially overlapping with ours are studied.

\section{${\bf CP}^n$ as a classical phase space}\label{cipienne}

We will first consider a classical mechanics whose phase space is complex, 
projective $n$--dimensional space ${\bf CP}^n$.
The following properties are well known \cite{KN}.

Let $Z^1,\ldots, Z^{n+1}$ denote homogeneous coordinates on ${\bf CP}^n$. 
The chart defined by $Z^k\neq 0$ covers one copy of the open set 
${\cal U}_k={\bf C}^n$. On the latter we have the holomorphic coordinates 
$z^j_{(k)}=Z^j/Z^k$, $j\neq k$; there are $n+1$ such coordinate charts. 
${\bf CP}^n$ is a K\"ahler manifold with respect 
to the Fubini--Study metric. On the chart $({\cal U}_k, z_{(k)})$ the K\"ahler 
potential reads
\begin{equation}
K(z^j_{(k)}, {\bar z}^j_{(k)})=
\log{\left(1 + \sum_{j=1}^n z^j_{(k)} {\bar z}^j_{(k)}\right)}.
\label{fubst}
\end{equation} 
The singular homology ring $H_*\left({\bf CP}^n, {\bf Z}\right)$ contains 
the nonzero subgroups 
\begin{equation}
H_{2k}\left({\bf CP}^n, {\bf Z}\right)={\bf Z}, \qquad 
k=0,1,\ldots, n,
\label{oncero}
\end{equation}
while 
\begin{equation}
H_{2k+1}\left({\bf CP}^n, {\bf Z}\right)=0, \qquad 
k=0,1,\ldots, n-1.
\label{oncerox}
\end{equation}
We have ${\bf CP}^{n}={\bf C}^n\cup {\bf CP}^{n-1}$, with ${\bf CP}^{n-1}$ a hyperplane 
at infinity. Topologically, ${\bf CP}^{n}$ is obtained by attaching a (real) $2n$--dimensional 
cell to ${\bf CP}^{n-1}$. ${\bf CP}^n$ is simply connected,
\begin{equation}
\pi_1\left({\bf CP}^n\right)=0,
\label{grfund}
\end{equation}
it is compact, and inherits its complex structure from that on ${\bf C}^{n+1}$. 

Let $\tau^{-1}$ denote the {\it tautological bundle}\/ on ${\bf CP}^n$. We recall that 
$\tau^{-1}$ is defined as the subbundle of the trivial bundle ${\bf CP}^n\times 
{\bf C}^{n+1}$ whose fibre at $p\in {\bf CP}^n$ is the line in ${\bf C}^{n+1}$ 
represented by $p$. Then $\tau^{-1}$ is a holomorphic line bundle over ${\bf CP}^n$. 
Its dual, denoted $\tau$, is called the {\it hyperplane bundle}. 
For any $l\in {\bf Z}$, the $l$--th power $\tau ^l$ is also a holomorphic line bundle 
over ${\bf CP}^n$. In fact every holomorphic line bundle 
$L$ over ${\bf CP}^n$ is isomorphic to $\tau ^l$ for some $l\in {\bf Z}$;
this integer is the first Chern class of $L$.

\section{The quantum line bundle}\label{qlb}

In the framework of geometric quantisation \cite{GEOMQUANT}
it is customary to consider the case when ${\cal C}$ is a compact 
K\"ahler manifold. In this context one introduces the notion of a quantisable, 
compact, K\"ahler phase space ${\cal C}$, of which ${\bf CP}^n$ is an 
example. This means that there exists a {\it quantum line bundle}\/ 
$({\cal L}, g, \nabla)$ on ${\cal C}$, where ${\cal L}$ is a holomorphic line bundle, 
$g$ a Hermitian metric on ${\cal L}$, and $\nabla$ a covariant derivative compatible 
with the complex structure and $g$. Furthermore, the curvature 
$F$ of $\nabla$ and the symplectic 2--form $\omega$ are required to satisfy
\begin{equation}
F=-2\pi {\rm i} \omega.
\label{quanxti}
\end{equation}
It turns out that quantisable, compact K\"ahler manifolds are projective 
algebraic manifolds and viceversa \cite{SCHLICHENMAIER}. After introducing 
a polarisation, the Hilbert space of quantum states is given by 
the global holomorphic sections of ${\cal L}$.

Recalling that, on ${\bf CP}^n$, ${\cal L}$ is isomorphic to $\tau^l$ for some 
$l\in {\bf Z}$, let ${\cal O}(l)$ denote the sheaf of holomorphic sections  
of ${\cal L}$ over ${\bf CP}^n$. The vector space of holomorphic sections of 
${\cal L}=\tau^l$ is the sheaf cohomology space $H^0({\bf CP}^n, {\cal O}(l))$.
The latter is zero for $l<0$, while for $l\geq 0$ it can be canonically 
identified with the set of homogeneous polynomials of degree $l$ on ${\bf C}^{n+1}$.
This set is a vector space of dimension $\left({n+l\atop 
n}\right)$:
\begin{equation}
{\rm dim}\,H^0({\bf CP}^n, {\cal O}(l))=\left({n+l\atop n}\right).
\label{jjoo}
\end{equation}
We will give a quantum--mechanical derivation of eqn. (\ref{jjoo}) in 
section \ref{esstqmb}.

Equivalence classes of holomorphic line bundles over a complex manifold 
${\cal C}$ are classified by the Picard group ${\rm Pic}\,({\cal C})$. 
The latter is defined \cite{GFH} as the sheaf cohomology group 
$H^1_{\rm sheaf}({\cal C}, {\cal O}^*)$,
where ${\cal O}^*$ is the sheaf of nonzero holomorphic functions on ${\cal C}$.
When ${\cal C}={\bf CP}^n$ things simplify because the above
sheaf cohomology group is in fact isomorphic to a singular homology group,
\begin{equation}
H^1_{\rm sheaf}({\bf CP}^n, {\cal O}^*)=H^2_{\rm sing}({\bf CP}^n, {\bf Z}),
\label{tsmplif}
\end{equation}
and the latter is given in eqn. (\ref{oncero}). Thus
\begin{equation}
{\rm Pic}\,({\bf CP}^n)={\bf Z}.
\label{athh}
\end{equation}
The zero class corresponds to the trivial line bundle;
all other classes correspond to nontrivial line bundles. 
As the equivalence class of ${\cal L}$ varies, 
so does the space ${\cal H}$ of its holomorphic sections vary.

\section{Quantum Hilbert--space bundles over ${\bf CP}^n$}\label{esstqmb}

In order to quantise ${\bf CP}^n$ we will construct a family of vector bundles 
over ${\bf CP}^n$, all of which will have a Hilbert space ${\cal H}$ as fibre.
We will analyse such bundles, that we will call {\it quantum 
Hilbert--space bundles},\/ or just ${\cal QH}$--bundles for short. Our aim 
is to demonstrate that there are different, nonequivalent choices for the 
${\cal QH}$--bundles, to classify them, and to study how the corresponding 
quantum mechanics varies with each choice.

Compactness of ${\bf CP}^n$ implies that, upon quantisation, the Hilbert space 
${\cal H}$ is finite--dimensional, and hence isomorphic 
to ${\bf C}^{N+1}$ for some $N$. This property follows from the fact that 
the number of quantum states grows monotonically with the symplectic volume 
of ${\cal C}$; the latter is finite when ${\cal C}$ is compact.
We are thus led to considering principal $U(N+1)$--bundles 
over ${\bf CP}^n$ and to their classification. Equivalently, we will consider 
the associated holomorphic vector bundles with fibre ${\bf C}^{N+1}$.
The corresponding projective bundles are ${\bf CP}^N$--bundles and principal $PU(N)$--bundles. 
Each choice of a different equivalence class of bundles will give rise to a different 
quantisation. 

So far we have left $N$ undetermined. In order to fix it we first 
pick the symplectic volume form $\omega^n$ on ${\bf CP}^n$ such that 
\begin{equation}
\int_{{\bf CP}^n}\omega^n=n+1.
\label{boludo}
\end{equation}
Next we set $N=n$, so ${\rm dim}\,{\cal H}=n+1$. This normalisation corresponds 
to 1 quantum state per unit of symplectic volume on ${\bf CP}^n$. Thus, {\it e.g.}, 
when $n=1$ we have the Riemann sphere ${\bf CP}^1$ and ${\cal H}={\bf C}^2$. 
The latter is the Hilbert space of a spin $s=1/2$ system, and the counting 
of states is correct. There are a number of further advantages to this normalisation. 
In fact eqn. (\ref{boludo}) is more than just a normalisation, in the sense that the 
dependence of the right--hand side on $n$ is determined by physical consistency arguments. 
This will be explained in section \ref{xcompt}.
Normalisation arguments can enter eqn. (\ref{boludo}) only through overall numerical 
factors such as $2\pi$, i$\hbar$, or similar. It is these latter factors that we fix by hand 
in eqn. (\ref{boludo}). 

The right--hand of our normalisation (\ref{boludo}) differs from 
that corresponding to eqn. (\ref{quanxti}). Up to numerical factors such as 
$2\pi$, ${\rm i}\hbar$, etc, it is standard  to set $\int_{{\bf CP}^n} F^n = 
n$ \cite{LIBAZCA}. There is also an alternative normalisation developed
in ref. \cite{BOYA}. However we will find our normalisation (\ref{boludo}) 
more convenient.

\subsection{Computation of ${\rm dim}\,H^0({\bf CP}^n, {\cal O}(1))$}\label{xcompt}

Next we present a quantum--mechanical computation of
${\rm dim}\,H^0({\bf CP}^n, {\cal O}(1))$ without resorting to sheaf 
cohomology. That is, we compute ${\rm dim}\, {\cal H}$ when $l=1$
and prove that it coincides with the right--hand side of eqn. (\ref{boludo}).
The case $l>1$ will be treated in section \ref{xcomptx}.

Starting with ${\cal C}={\bf CP}^{0}$, {\it i.e.}, a point $p$ as classical phase space, 
the space of quantum rays must also reduce to a point. Then the corresponding Hilbert space 
is ${\cal H}_1={\bf C}$. The only state in ${\cal H}_1$ is the vacuum $|0\rangle_{l=1}$, 
henceforth denoted $|0\rangle$ for brevity. 

Next we pass from ${\cal C}={\bf CP}^0$ to ${\cal C}={\bf CP}^1$. 
Regard $p$, henceforth denoted $p_1$, as the {\it point at infinity}\/ 
with respect to a coordinate chart $({\cal U}_1, z_{(1)})$ on ${\bf CP}^1$ that does not 
contain $p_1$. This chart is biholomorphic to ${\bf C}$ and supports a representation 
of the Heisenberg algebra in terms of creation and annihilation operators $A^{\dagger}(1)$,  
$A(1)$. This process adds the new state $A^{\dagger}(1)|0(1)\rangle$ to the spectrum. 
The new Hilbert space ${\cal H}_2={\bf C}^2$ is the linear span of $|0(1)\rangle$ 
and $A^{\dagger}(1)|0(1)\rangle$.

On ${\bf CP}^1$ we have the charts $({\cal U}_1, z_{(1)})$ and $({\cal U}_2, z_{(2)})$. 
Point $p_1$ is at infinity with respect to $({\cal U}_1, z_{(1)})$, while it 
belongs to $({\cal U}_2, z_{(2)})$. Similarly, the point at infinity with respect to 
$({\cal U}_2, z_{(2)})$, call it $p_2$, belongs to $({\cal U}_1, z_{(1)})$ but not to 
$({\cal U}_2, z_{(2)})$. Above we have proved that the Hilbert--space bundle ${\cal QH}_2$ 
has a fibre ${\cal H}_2={\bf C}^2$ which, on the chart ${\cal U}_1$, is the linear span 
of $|0(1)\rangle$ and $A^{\dagger}(1)|0(1)\rangle$. On the chart ${\cal U}_2$, the 
fibre is the linear span of $|0(2)\rangle$ and $A^{\dagger}(2)|0(2)\rangle$, 
$A^{\dagger}(2)$ being the creation operator on ${\cal U}_2$. On the common overlap 
${\cal U}_1\cap {\cal U}_2$, the coordinate transformation between $z_{(1)}$ and 
$z_{(2)}$ is holomorphic. This implies that, on ${\cal U}_1\cap {\cal U}_2$,
the fibre ${\bf C}^2$ can be taken in either of two equivalent ways: either 
as the linear span of $|0(1)\rangle$ and $A^{\dagger}(1)|0(1)\rangle$, or as that of
$|0(2)\rangle$ and $A^{\dagger}(2)|0(2)\rangle$. 

The general construction is now clear. Topologically we have ${\bf CP}^{n}={\bf 
C}^n\cup {\bf CP}^{n-1}$, with ${\bf CP}^{n-1}$ a hyperplane at infinity, 
but we also need to describe the coordinate charts and their overlaps.
There are coordinate charts $({\cal U}_j, z_{(j)})$, $j=1, \ldots, n+1$ 
and nonempty $f$--fold overlaps $\cap_{j=1}^f {\cal U}_j$ for $f=2,3,\ldots, n+1$. 
Each chart $({\cal U}_j, z_{(j)})$ is biholomorphic with ${\bf C}^n$ and has
a ${\bf CP}^{n-1}$--hyperplane at infinity; the latter is charted by the 
remaining charts $({\cal U}_k, z_{(k)})$, $k\neq j$.
Over $({\cal U}_j, z_{(j)})$ the Hilbert--space bundle ${\cal QH}_{n+1}$ has a fibre 
${\cal H}_{n+1}={\bf C}^{n+1}$ spanned by 
\begin{equation}
|0(j)\rangle,\qquad  A_i^{\dagger}(j) |0(j)\rangle, \qquad i=1,2,\ldots, n. 
\label{pann}
\end{equation}
Analyticity arguments similar to those above prove that, on every nonempty 
$f$--fold overlap $\cap_{j=1}^f {\cal U}_j$, the fibre ${\bf C}^{n+1}$ can 
be taken in $f$ different, but equivalent ways, as the linear span of 
$|0(j)\rangle$ and $A_i^{\dagger}(j) |0(j)\rangle$,  $i=1,2,\ldots, n$, for 
every choice of $j=1,\ldots, f$.

A complete description of this bundle requires the specification of the 
transition functions. We take the excited states $A_i^{\dagger}(j) |0(j)\rangle$
to transform according to the jacobian matrices $t(T{\bf CP}^n)$ corresponding to 
coordinate changes on ${\bf CP}^n$, while the vacuum $|0\rangle$ will 
transform with the transition functions $t({\tau})$ of the line bundle $\tau$.
Thus the complete transition functions are the direct sum
\begin{equation}
t({\cal QH}({\bf CP}^n))=t(T{\bf CP}^n)\oplus t({\tau}),
\label{jodetelabastidacabron}
\end{equation}
and the ${\cal QH}$--bundle itself decomposes as the direct sum of a 
holomorphic line bundle $N({\bf CP}^n)=\tau$, plus the holomorphic 
trangent bundle $T({\bf CP}^n)$,
\begin{equation}
{\cal QH}({\bf CP}^n)=T({\bf CP}^n) \oplus N({\bf CP}^n).
\label{labastidaputa}
\end{equation}
It follows that tangent vectors to ${\bf CP}^n$ are quantum states 
in (the defining representation of) Hilbert space. In eqn. (\ref{pann}) we 
have given a basis for these states in terms of creation operators acting 
on the vacuum $|0\rangle$. The latter can be regarded as the basis vector 
for the fibre ${\bf C}$ of the line bundle $N({\bf CP}^n)$.

\subsection{Representations}\label{wrepp}

The $(n+1)$--dimensional Hilbert space of eqn. (\ref{pann}) may be regarded as a kind of 
{\it defining representation}, in the sense of the representation theory 
of $SU(n+1)$. The latter is the structure group of the bundle (\ref{labastidaputa}).
Comparing our results with those of section \ref{qlb} we conclude 
that ${\cal L}=\tau$, because $l=1$. 
This is the smallest value of $l$ that produces a nontrivial 
${\cal H}$, as eqn. (\ref{jjoo}) gives a 1--dimensional Hilbert space when $l=0$. 
So our ${\cal H}$  spans an $(n+1)$--dimensional representation of $SU(n+1)$, that we can 
identify with the defining representation. There is some ambiguity here since the 
dual of the defining representation of $SU(n+1)$ is also $(n+1)$--dimensional. 
This ambiguity is resolved by convening that the latter is generated by the 
holomorphic sections of the {\it dual}\/ quantum line bundle ${\cal L}^*=\tau^{-1}$.
On the chart ${\cal U}_j$, $j=1,\ldots, n+1$, 
the dual of the defining representation is the linear span of the covectors
\begin{equation}
\langle (j)0|,\qquad  \langle (j)0|A_i(j), \qquad i=1,2,\ldots, n. 
\label{pannx}
\end{equation}
Taking higher representations is equivalent to considering the principal
$SU(n+1)$--bundle (associated with the vector ${\bf C}^{n+1}$--bundle) in a 
representation higher than the defining one. We will see next
that this corresponds to having $l>1$ in our choice of the line bundle $\tau^l$.

\subsection{Computation of ${\rm dim}\,H^0({\bf CP}^n, {\cal O}(l))$}\label{xcomptx}

We extend now our quantum--mechanical computation of ${\rm dim}\,H^0({\bf CP}^n, 
{\cal O}(l))$ to the case $l>1$. As in section \ref{xcompt}, we do not resort to 
sheaf cohomology. The values $l=0,1$ respectively correspond to the trivial and the 
defining representation of $SU(n+1)$. The restriction to nonnegative $l$ follows 
from our convention of assigning the defining representation to $\tau$ and 
its dual to $\tau^{-1}$. 
Higher values $l>1$ correspond to higher representations and can be accounted for as follows. 
We have 
\begin{equation}
{\bf CP}^{n+l}=SU(n+l+1)/\left(SU(n+l)\times U(1)\right),
\label{xxaa}
\end{equation}
where now $SU(n+l+1)$ and $SU(n+l)$ act on ${\bf C}^{n+l+1}$. Now
$SU(n+l)$ admits $\left({n+l}\atop n\right)$--dimensional representations 
(Young tableaux with a single column of $n$ boxes) that, by restriction, 
are also representations of $SU(n+1)$. Letting $l>1$ vary for fixed $n$,
this reproduces the dimension of eqn. (\ref{jjoo}).

By itself, the existence of $SU(n+1)$ representations with the dimension
of eqn. (\ref{jjoo}) does not prove that, picking $l>1$, the corresponding 
quantum states lie in those $\left({n+l}\atop n\right)$--dimensional representations.
We have to prove that no other value of the dimension fits the given data.
In order to prove it the idea is, roughly speaking, that a value of $l>1$ on 
${\bf CP}^n$ can be traded for $l'=1$ on ${\bf CP}^{n+l}$. That is, an $SU(n+1)$ 
representation higher than the defining one can be traded for the defining 
representation of $SU(n+l+1)$. In this way the ${\cal QH}$--bundle on ${\bf 
CP}^n$ with the Picard class $l'=l$ equals the ${\cal QH}$--bundle on ${\bf 
CP}^{n+l}$ with the Picard class $l'=1$.
On the latter we have $n+l$ excited states ({\it i.e.}, other than the 
vacuum), one for each complex dimension of ${\bf CP}^{n+l}$. We can sort 
them into unordered sets of $n$, which is the number of excited states on 
${\bf CP}^n$, in $\left({n+l}\atop n\right)$ different ways. This selects
a specific dimension for the $SU(n+1)$ representations and rules out the rest.
More precisely, it is only when $n>1$ that some representations are ruled 
out. When $n=1$, {\it i.e.} for $SU(2)$, all representations are allowed, 
since their dimension is $l+1=\left({1+l}\atop 1\right)$. However already for 
$SU(3)$ some representations are thrown out. The number $\left({2+l}\atop 2\right)$
matches the dimension $d(p,q)=(p+1)(q+1)(p+q+2)/2$ of the $(p,q)$ 
irreducible representation if $p=0$ and $l=q$ or $q=0$ and $l=p$,
but arbitrary values of $(p,q)$ are in general not allowed.

To complete our reasoning we have to prove that the quantum line bundle 
${\cal L}=\tau$ on ${\bf CP}^{n+l}$ descends to ${\bf CP}^{n}$ as the $l$--th power 
$\tau^l$. For this we resort to the natural embedding of ${\bf CP}^{n}$ 
into ${\bf CP}^{n+l}$.
Let $({\cal U}_{1}, z_{(1)})$, $\ldots$, $({\cal U}_{n+1}, z_{(n+1)})$ be the 
coordinate charts on ${\bf CP}^{n}$ described in section \ref{cipienne},
and let $(\tilde{\cal U}_{1}, \tilde z_{(1)})$, $\ldots$, $(\tilde{\cal U}_{n+1}, 
\tilde z_{(n+1)})$, $(\tilde{\cal U}_{n+2}, \tilde z_{(n+2)})$, $\ldots$, 
$(\tilde{\cal U}_{n+l+1}, \tilde z_{(n+l+1)})$ 
be charts on ${\bf CP}^{n+l}$ relative to this embedding. This means that the first 
$n+1$ charts on ${\bf CP}^{n+l}$, duly restricted, are also charts on ${\bf CP}^{n}$; 
in fact every chart on ${\bf CP}^n$ is contained $l$ times within ${\bf CP}^{n+l}$. 
Let $t_{jk}(\tau)$, with $j,k=1,\ldots, n+l+1$, be the transition function for $\tau$ 
on the overlap $\tilde{\cal U}_{j}\cap \tilde{\cal U}_{k}$ of ${\bf CP}^{n+l}$. 
In passing from $\tilde{\cal U}_{j}$ to $\tilde{\cal U}_{k}$,
points on the fibre are acted on by $t_{jk}(\tau)$. Due to our choice of embedding, 
the overlap $\tilde{\cal U}_{j}\cap \tilde{\cal U}_{k}$ on ${\bf CP}^{n+l}$ contains 
$l$ copies of the overlap ${\cal U}_{j}\cap {\cal U}_{k}$ on ${\bf CP}^{n}$. 
Thus points on the fibre over ${\bf CP}^n$ are acted on by $(t_{jk}(\tau))^l$, 
where now $j,k$ are restricted to $1,\ldots, n+1$. This means that the line 
bundle on ${\bf CP}^n$ is $\tau^l$ as stated, and the vacuum $|0\rangle_{l'=l}$ 
on ${\bf CP}^n$ equals the vacuum $|0\rangle_{l'=1}$ on ${\bf CP}^{n+l}$. Hence 
there are on ${\bf CP}^n$ as many inequivalent vacua as there are elements in ${\bf 
Z}={\rm Pic}\,({\bf CP}^n)$ (remember that sign reversal $l\rightarrow -l$ 
within ${\rm Pic}\,({\bf CP}^n)$ is the operation of taking the dual 
representation, {\it i.e.}, $\tau\rightarrow \tau^{-1}$).

\subsection{Classification of ${\cal QH}$--bundles}\label{mmrrbb}

As a holomorphic line bundle, $N({\bf CP}^n)$ is isomorphic to $\tau^l$ for some 
$l\in {\rm Pic}\,({\bf CP}^n)$ $={\bf Z}$. Now the bundle $T({\bf CP}^n)
\oplus N({\bf CP}^n)$ has $SU(n+1)$ as its structure group, which we consider 
in the representation $\rho_l$ corresponding to the Picard class $l\in {\bf Z}$:
\begin{equation}
{\cal QH}_l({\bf CP}^n)=\rho_l(T({\bf CP}^n)) \oplus \tau^l, \qquad l\in {\bf Z}.
\label{adxx}
\end{equation} 
The above generalises eqn. (\ref{labastidaputa}) to the case $l>1$.
The importance of eqn. (\ref{adxx}) is that it classifies ${\cal QH}$--bundles 
over ${\bf CP}^n$: holomorphic equivalence classes of such bundles are in 1--to--1 
correspondence with the elements of ${\bf Z}={\rm Pic}\,({\bf CP}^n)$. The 
class $l=1$ corresponds to the defining representation of $SU(n+1)$,
\begin{equation}
{\cal QH}_{l=1}({\bf CP}^n)=T({\bf CP}^n)\oplus\tau,
\label{pemex}
\end{equation}
and $l=-1$ to its dual. 
The quantum Hilbert--space bundle over ${\bf CP}^n$ is generally nontrivial, 
although particular values of $l$ may render the direct sum (\ref{adxx}) trivial. 
The separate summands $T({\bf CP}^n)$ and $N({\bf CP}^n)$ are both nontrivial bundles. 
Nontriviality of $N({\bf CP}^n)$ means that, when $l\neq 0$, the state 
$|0\rangle$ transforms nontrivially (albeit as multiplication by a phase factor)
between different local trivialisations of the bundle. When $l=0$ the vacuum 
transforms trivially.

According to eqn. (\ref{adxx}), the transition functions $t({\cal QH}_l)$ for  
${\cal QH}_l$ decompose as a direct sum of two transition functions, 
one for $\rho_l(T({\bf CP}^n))$, another one for $\tau^l$:
\begin{equation}
t({\cal QH}_l({\bf CP}^n))=t(\rho_l(T{\bf CP}^n))\oplus t({\tau^l}).
\label{decc}
\end{equation}
If the transition functions for $\tau$ are $t(\tau)$, those for 
$\tau^l$ are $(t(\tau))^l$. On the other hand, the transition functions 
$t(\rho_l(T{\bf CP}^n))$ are the jacobian matrices (in representation 
$\rho_l$) corresponding to coordinate changes on ${\bf CP}^n$. 
Then all the  ${\cal QH}_l({\bf CP}^n)$--bundles of eqn. (\ref{adxx}) are nonflat 
because the tangent bundle $T({\bf CP}^n)$ itself is nonflat. Eqn. (\ref{decc})
generalises eqn. (\ref{jodetelabastidacabron}) to the case $l>1$.

\subsection{Diagonalisation of the projective Hamiltonian}\label{tnclmm}

Deleting from ${\bf CP}^n$ the ${\bf CP}^{n-1}$--hyperplane at infinity produces the noncompact 
space ${\bf C}^n$. The latter is the classical phase space of the $n$--dimensional harmonic 
oscillator (now no longer {\it projective}\/, but {\it linear}\/). The corresponding Hilbert space 
${\cal H}$ is infinite--dimensional because the symplectic volume of ${\bf C}^n$ is infinite.

The deletion of the hyperplane at infinity may also be understood from the 
viewpoint of the K\"ahler potential (\ref{fubst}) corresponding to the 
Fubini--Study metric. No longer being able to pass holomorphically 
from a point at finite distance to a point at infinity implies that, 
on the conjugate chart $({\cal U}_k, z_{(k)})$, the squared modulus 
$|z_{(k)}|^2$ is always small and we can Taylor--expand eqn. (\ref{fubst}) as
\begin{equation}
\log{\left(1 + \sum_{j=1}^n z^j_{(k)} {\bar z}^j_{(k)}\right)}\simeq \sum_{j=1}^n 
z^j_{(k)} {\bar z}^j_{(k)}.
\label{tayy}
\end{equation}
The right--hand side of eqn. (\ref{tayy}) is the K\"ahler potential for 
the usual Hermitean metric on ${\bf C}^n$. As such, $\sum_{j=1}^n z^j_{(k)} {\bar 
z}^j_{(k)}$ equals the classical Hamiltonian for the $n$--dimensional linear harmonic 
oscillator. Observers on this coordinate chart effectively see ${\bf C}^n$ 
as their classical phase space. The corresponding Hilbert space is the 
(closure of the) linear span of the states $|m_1,\ldots, m_n\rangle$, where 
\begin{equation}
H_{\rm lin}|m_1,\ldots, m_n\rangle = \sum_{j=1}^n 
\left(m_j + {1\over 2}\right)|m_1,\ldots, m_n\rangle,\qquad 
m_j=0,1,2,\ldots,
\label{oegg}
\end{equation}
and 
\begin{equation}
H_{\rm lin}=\sum_{j=1}^n\left(A^{\dagger}_j(k)A_j(k)+{1\over 2}\right)
\label{hachelin}
\end{equation} 
is the quantum Hamiltonian operator corresponding to the classical 
Hamiltonian function on the right--hand side of eqn. (\ref{tayy}). Then the stationary 
Schr\"odinger equation for the {\it projective}\/ oscillator reads
\begin{equation}
H_{\rm proj}|m_1,\ldots, m_n\rangle = \log\left(1+\sum_{j=1}^n 
\left(m_j + {1\over 2}\right)\right)|m_1,\ldots, m_n\rangle,
\label{oeggpp}
\end{equation}
where 
\begin{equation}
H_{\rm proj}=\log\left(1+\sum_{j=1}^n\left(A^{\dagger}_j(k)A_j(k)+
{1\over 2}\right)\right)
\label{hacheproj}
\end{equation}
is the quantum Hamiltonian operator corresponding to 
the classical Hamiltonian function on the left--hand side of eqn. (\ref{tayy}).

The same states $|m_1,\ldots, m_n\rangle$ that diagonalise $H_{\rm lin}$ also 
diagonalise $H_{\rm proj}$. However, eqns. (\ref{oegg})--(\ref{hacheproj}) 
above in fact only hold locally on the chart ${\cal U}_k$, which does not cover 
all of ${\bf CP}^n$. Bearing in mind that there is one hyperplane at infinity 
with respect to this chart, we conclude that the arguments of section \ref{xcompt}
apply in order to ensure that the projective oscillator only has $n$ excited states.
Then the occupation numbers $m_j$ are either all 0 (for the vacuum state) 
or all zero but for one of them, where $m_j=1$ (for the excited states), 
and ${\rm dim}\,{\cal H}=n+1$ as it should. Moreover, the eigenvalues of eqn. (\ref{oeggpp})
provide an alternative proof of the fact, demonstrated in section \ref{xcomptx}, 
that the Picard group class $l'=l>1$ on ${\bf CP}^n$ can be 
traded for $l'=1$ on ${\bf CP}^{n+l}$.

\section{${\bf CP}({\cal H})$ as a classical phase space}\label{innter}

Realise ${\cal H}$ as the space of infinite sequences of complex numbers $Z^1, Z^2, \ldots$ that 
are square--summable, $\sum_{j=1}^{\infty}\vert Z^j\vert ^2 < \infty$. 
The $Z^j$ provide a set of holomorphic coordinates on ${\cal H}$. 
The space of rays ${\bf CP}({\cal H})$ is
\begin{equation}
{\bf CP}({\cal H})=({\cal H}-\{0\})/({\bf R}^+\times U(1)).
\label{cph}
\end{equation}

The $Z^j$ provide a set of {\it projective}\/ coordinates on ${\bf CP}({\cal H})$.
Now assume that $Z^k\neq 0$, and define $z^j_{(k)}= Z^j/Z^k$ for $j\neq k$. Then 
$\sum_{j\neq k}^{\infty}\vert z^j_{(k)}\vert ^2 < \infty$ for every fixed 
value of $k$. As $j\neq k$ varies, these $z^j_{(k)}$ cover one copy of ${\cal H}$ 
that we denote by ${\cal U}_k$. The open set ${\cal U}_k$, endowed with 
the coordinate functions $z^j_{(k)}$, $j=1,2, \ldots \check k, \ldots$, 
where a check over an index indicates omission, provides a holomorphic coordinate chart 
on ${\bf CP}({\cal H})$ for every fixed $k$.
A holomorphic atlas is obtained as the collection of all pairs  $({\cal U}_k, 
z_{(k)})$, for $k=1, 2, \ldots$ There are nonempty $f$--fold overlaps $\cap_{m=1}^f {\cal U}_m$ 
for all values of $f=1, 2, \ldots$ When $f=2$, tangent vectors transform according to an 
(infinite--dimensional) jacobian matrix.

${\bf CP}({\cal H})$ is a K\"ahler manifold. On the coordinate chart $({\cal 
U}_k, z_{(k)})$, the K\"ahler potential reads
\begin{equation}
K(z_{(k)}, {\bar z}_{(k)})=
\log{\left(1 + \sum_{j\neq k}^{\infty} z^j_{(k)} {\bar z}^j_{(k)}\right)},
\label{fubstx}
\end{equation} 
and the corresponding metric ${\rm d}s^2_K$ reads on this chart
\begin{equation}
{\rm d}s^2_K = \sum_{m,n\neq k}^{\infty}{\partial^2 K(z_{(k)}, {\bar z}_{(k)})\over
\partial z_{(k)}^m \partial {\bar z}_{(k)}^n}\, {\rm d}z_{(k)}^m{\rm d}{\bar z}_{(k)}^n.
\label{rrdd}
\end{equation}
Being infinite--dimensional, ${\bf CP}({\cal H})$ is noncompact. It is simply connected:
\begin{equation}
\pi_1\left({\bf CP}({\cal H})\right)=0.
\label{cphx}
\end{equation}
Its Picard group is the group of integers:
\begin{equation}
{\rm Pic}\, ({\bf CP}({\cal H}))={\bf Z}.
\label{ppic}
\end{equation}
It has trivial homology in odd real dimension,
\begin{equation}
H_{2k+1}\left({\bf CP}({\cal H}), {\bf Z}\right)=0,\qquad k=0,1,\ldots,
\label{cotri}
\end{equation}
while it is nontrivial in even dimension,
\begin{equation}
H_{2k}\left({\bf CP}({\cal H}), {\bf Z}\right)={\bf Z}, \qquad k=0,1,\ldots
\label{pelotudo}
\end{equation}

\section{Quantum Hilbert--space bundles over ${\bf CP}({\cal H})$}\label{cepehache}

By eqn. (\ref{ppic}), for each integer $l\in{\bf Z}$ there exists one equivalence class 
$N_l({\bf CP}({\cal H}))$ of holomorphic lines bundles over ${\bf CP}({\cal H})$. 
For $l\neq 0$ this bundle is nontrivial; its fibre ${\bf C}$ is generated by the vacuum state 
$\vert 0\rangle_l$. Let $A^{\dagger}_j(k)$, $A_j(k)$, $j\neq k$, 
be creation and annihilation operators on the chart ${\cal U}_k$, for $k$ fixed. 
We can now construct the ${\cal QH}_l$--bundle over ${\bf CP}({\cal H})$. To this end 
we will describe the fibre over each coordinate chart ${\cal U}_k$, plus the transition 
functions on the 2--fold overlaps ${\cal U}_k\cap {\cal U}_m$, for all $k\neq m$. 

The Hilbert--space fibre over ${\cal U}_k$ is ${\cal H}$ itself, the latter being the 
${\bf C}$--linear span of the infinite set of linearly independent vectors 
\begin{equation}
\vert 0(k)\rangle_l,\qquad A^{\dagger}_j(k)\vert 0(k)\rangle_l,\qquad 
j=1, 2, \ldots, \check k, \ldots
\label{nnumm}
\end{equation}
Reasoning as in section \ref{esstqmb} one proves that, 
on the 2--fold overlaps ${\cal U}_k\cap {\cal U}_m$, the fibre ${\cal H}$ can be chosen 
in either of two equivalent ways.  ${\cal H}$ is either the ${\bf C}$--linear span 
of the vectors $\vert 0(k)\rangle_l$, $A^{\dagger}_j(k)\vert 0(k)\rangle_l$, 
for $j=1, 2, \ldots, \check k, \ldots$, or the ${\bf C}$--linear span of the vectors 
$\vert 0(m)\rangle_l$, $A^{\dagger}_j(m)\vert 0(m)\rangle_l$, for $j=1, 2, \ldots, \check m, \ldots$

As in section \ref{esstqmb} we have that the vacuum $\vert 0(k)\rangle_l$ is the 
fibrewise generator of a holomorphic line bundle $N_l({\bf CP}({\cal H}))$.
Its excitations $A^{\dagger}_j(k)\vert 0(k)\rangle_l$ are tangent vectors to ${\bf CP}({\cal H})$ 
on the chart ${\cal U}_k$, and thus transition functions are the sum of
two parts. One is a phase factor accounting for the transformation of
$\vert 0(k)\rangle_l$; the other one is a jacobian matrix.
The complete ${\cal QH}_l$--bundle splits as
\begin{equation}
{\cal QH}_l({\bf CP}({\cal H})) = T({\bf CP}({\cal H}))\oplus  N_l({\bf CP}({\cal H})).
\label{quhache}
\end{equation}

\section{Quantum Hilbert--space bundles over ${\cal C}$}\label{imbbdd}

Next we present a summary, drawn from ref. \cite{CPINF}, 
on how to holomorphically embed a noncompact ${\cal C}$ within ${\bf CP}({\cal H})$. 
This procedure is applied in section \ref{qqcccepeh} in order to quantise ${\cal C}$.

\subsection{The Bergman metric on ${\cal C}$}\label{bemet}

Denote by ${\cal F}$ the set of holomorphic, square--integrable $n$--forms on ${\cal C}$.
${\cal F}$ is a separable, complex Hilbert space (finite--dimensional when ${\cal C}$ is compact).
Let $h_1, h_2, \ldots $ denote a complete orthonormal basis for ${\cal F}$, and let $z$ be (local) 
holomorphic coordinates on ${\cal C}$. Then
\begin{equation}
{\cal K}(z, \bar w)=\sum_{j=1}^{\infty}h_j(z)\wedge\bar h_j(\bar w)
\label{kaka}
\end{equation}
is a holomorphic $2n$--form on ${\cal C}\times\bar {\cal C}$, where $\bar {\cal C}$
is complex manifold conjugate to ${\cal C}$. The form ${\cal K}(z,\bar w)$ is 
independent of the choice of an orthonormal basis for ${\cal F}$; it is called 
the {\it kernel form}\/ of ${\cal C}$. If $\bar z$ is the point of $\bar {\cal C}$ 
corresponding to a point $z\in{\cal C}$, the set of pairs $(z,\bar 
z)\in {\cal C}\times \bar {\cal C}$ is naturally identified with ${\cal 
M}$. In this way ${\cal K}(z, \bar z)$ can be considered as a $2n$--form on 
${\cal C}$. One can prove that ${\cal K}(z, \bar z)$ is invariant under the group 
of holomorphic transformations of ${\cal C}$.

Next assume that, given any point $z\in{\cal C}$, there exists an 
$f\in{\cal F}$ such that $f(z)\neq 0$. That is, the kernel form 
${\cal K}(z,\bar z)$ of ${\cal C}$ is everywhere nonzero on ${\cal C}$:
\begin{equation}
{\cal K}(z,\bar z)\neq 0,\qquad \forall z\in{\cal C}.
\label{knz}
\end{equation}
Let us write, in local holomorphic coordinates $z^j$ on ${\cal C}$,  $j=1,\ldots, n$,
\begin{equation}
{\cal K}(z,\bar z)={\bf k}(z, \bar z)\,{\rm d}z^1\wedge\ldots\wedge{\rm d}z^n\wedge{\rm d}\bar 
z^1\wedge\ldots\wedge{\rm d}\bar z^n,
\label{kicco}
\end{equation}
for a certain everywhere nonzero function ${\bf k}(z, \bar z)$. 
Define a hermitean form ${\rm d}s^2_B$
\begin{equation}
{\rm d}s^2_B=\sum_{j,k=1}^n {\partial^2 \log {\bf k}\over \partial z^j \bar z^k}
\,{\rm d}z^j {\rm d}\bar z^k.
\label{bbmmff}
\end{equation}
One can prove that ${\rm d}s^2_B$ is independent of the choice of coordinates 
on ${\cal C}$. Moreover, it is positive semidefinite and invariant under 
the holomorphic transformations of ${\cal C}$.

Let us make the additional assumption that ${\cal C}$ is such that ${\rm d}s^2_B$ 
is positive definite,
\begin{equation}
{\rm d}s^2_B>0.
\label{posdef}
\end{equation}
Then ${\rm d}s^2_B$ defines a (K\"ahler) metric  called the {\it Bergman metric} 
on ${\cal C}$ \cite{PERELOMOV}.

\subsection{Embedding ${\cal C}$ within ${\bf CP}({\cal H})$}\label{fedemb}

Let ${\cal H}$ be the Hilbert space dual to ${\cal F}$. Given $f\in {\cal F}$, let its expansion 
in local coordinates be
\begin{equation}
f={\bf f}\,{\rm d}z^1\wedge\ldots\wedge{\rm d}z^n,
\label{ppxx}
\end{equation}
for a certain function ${\bf f}$. Let $\iota'$ denote the mapping that sends $z\in {\cal C}$ 
into $\iota'(z)\in{\cal H}$ defined by
\begin{equation}
\langle\iota'(z)|f\rangle = {\bf f}(z).
\label{hhaa}
\end{equation}
Then $\iota'(z)\neq 0$ for all $z\in{\cal C}$ if and only if property (\ref{knz}) holds. 
Assuming that the latter is satisfied, and denoting by $p'$ the natural projection from 
${\cal H}-\{0\}$ onto ${\bf CP}({\cal H})$, the composite map $\iota = p'\circ\iota'$ 
\begin{equation}
\iota\colon{\cal C}\rightarrow{\bf CP}({\cal H})
\label{ccmmppp}
\end{equation}
is well defined on ${\cal C}$, independent of the coordinates, and holomorphic. 

One can prove the following results. When property (\ref{knz}) is true, the 
quadratic differential form ${\rm d}s^2_B$ of eqn. (\ref{bbmmff}) is the pullback, 
by $\iota$, of the canonical K\"ahler metric ${\rm d}s^2_K$ of eqn. (\ref{rrdd}):
\begin{equation}
{\rm d}s^2_B = \iota^*({\rm d}s^2_K).
\label{ppbbck}
\end{equation}
Moreover, the differential of $\iota$ is nonsingular at every point 
of ${\cal C}$ if and only if property (\ref{posdef}) is satisfied. 
These two results give us a geometric interpretation of the Bergman metric. 
Namely, if properties (\ref{knz}) and (\ref{posdef}) hold, 
then $\iota$ is an isometric immersion of ${\cal C}$ into ${\bf CP}({\cal H})$.

The map $\iota$ is locally one--to--one in the sense that every point of ${\cal C}$ 
has a neighbourhood that is mapped injectively into ${\bf CP}({\cal H})$. 
However, $\iota$ is not necessarily injective in the large. Conditions can 
be found that ensure injectivity of $\iota$ in the large. Assume 
that, if $z$, $z'$ are any two distinct points of ${\cal C}$, an 
$f\in{\cal F}$ can be found such that
\begin{equation}
f(z)\neq 0,\qquad f(z')=0.
\label{cdtn}
\end{equation}
Then $\iota$ is injective. Therefore, if ${\cal C}$ satisfies assumptions (\ref{knz}), 
(\ref{posdef}) and (\ref{cdtn}), it can be holomorphically and isometrically embedded 
into ${\bf CP}({\cal H})$.

\subsection{Quantisation of ${\cal C}$ as a submanifold of ${\bf CP}({\cal H})$}
\label{qqcccepeh}

Finally we quantise a noncompact ${\cal C}$ with infinite symplectic volume,
\begin{equation}
\int_{\cal C}\omega^n = \infty,
\label{isxvcc}
\end{equation}
so ${\cal H}$ will be infinite--dimensional. On the other hand,
${\cal C}$ admits only $n$ linearly independent, holomorphic tangent 
vectors, so the technique of section \ref{esstqmb} must be modified.

We need an infinite--dimensional ${\cal QH}$--bundle over ${\cal C}$. 
For this purpose we assume embedding ${\cal C}$ holomorphically 
and injectively within ${\bf CP}({\cal H})$ as in eqn. (\ref{ccmmppp}). 
Then the bundle ${\cal QH}_l({\bf CP}({\cal H}))$ of eqn. (\ref{quhache}) can 
be pulled back to ${\cal C}$ by the embedding $\iota$. We take this {\it to define}\/
the bundle ${\cal QH}_l({\cal C})$:
\begin{equation}
{\cal QH}_l({\cal C}) = \iota^* {\cal QH}_l({\bf CP}({\cal H})).
\label{ppbbac}
\end{equation}
Even if ${\cal QH}_l({\bf CP}({\cal H}))$ were trivial (which it is not for 
$l\neq 0$), it might contain nonflat (hence nontrivial) 
subbundles, thus allowing for nontrivial dualities.

A detailed analysis of ${\cal QH}_l({\cal C})$ requires specifying ${\cal C}$ explicitly.
However some properties can be stated in general. Thus, {\it e.g.},
the kernel form is the quantum--mechanical propagator. 
On ${\bf C}^n$ it reads 
\begin{equation}
{\cal K}_{{\bf C}^n}(z, \bar z) = N \,{\rm exp}\left({\rm i}\sum_{j=1}^n\bar z^j z^j\right)
{\rm d}z^1\wedge\ldots\wedge{\rm d}z^n\wedge{\rm d}\bar z^1\wedge\ldots\wedge{\rm d}\bar z^n,
\label{ceene}
\end{equation}
where $N$ is some normalisation. The Bergman metric (\ref{bbmmff}) derived from this kernel 
is the standard Hermitean metric on ${\bf C}^n$. The embedding $\iota$ naturally relates physical 
information (the propagator) and geometric information (the metric on ${\cal C}$). In retrospective, 
this justifies our quantisation of ${\cal C}$ by embedding it within ${\bf CP}({\cal H})$. 

\section{Summary}\label{dsskk}

Our analysis has dealt primarily with the case when ${\cal C}={\bf CP}^n$. 
In section \ref{qlb} we have recalled some well--known facts from 
geometric quantisation. They concern the dimension of the space of holomorphic 
sections of the quantum line bundle on a compact, quantisable K\"ahler 
manifold. This dimension has been rederived in section \ref{esstqmb} using purely 
quantum--mechanical arguments, by constructing the Hilbert--space bundle 
of quantum states over ${\bf CP}^n$. For brevity, the following summary 
deals only with the case when the Hilbert space is ${\bf C}^{n+1}$
(see sections \ref{wrepp}, \ref{xcomptx} for the general case).
The fibre ${\bf C}^{n+1}$ over a given coordinate chart 
on ${\bf CP}^n$ is spanned by the vacuum state $|0(j)\rangle_l$, 
plus $n$ states $A^{\dagger}_j|0(j)\rangle_l$, $j=1,\ldots, n$, 
obtained by the action of creation operators. We have identified 
the transition functions of this bundle as jacobian matrices plus a phase factor. 
The jacobian matrices account for the transformation (under coordinate changes on 
${\bf CP}^n$) of the states $A^{\dagger}_j|0(j)\rangle_l$, while the phase factor 
corresponds to $|0(j)\rangle_l$. This means that all quantum states (except the vacuum) 
are tangent vectors to ${\bf CP}^n$. In this way the Hilbert--space bundle over 
${\bf CP}^n$ splits as the direct sum of two holomorphic vector bundles: 
the tangent bundle $T({\bf CP}^n)$, plus a line bundle $N({\bf CP}^n)$ 
whose fibrewise generator is the vacuum. 

All complex manifolds admit a Hermitian metric, so having tangent vectors 
as quantum states suggests using the Hermitian connection and the corresponding 
curvature tensor to measure flatness. Now $T({\bf CP}^n)$ is nonflat,
so it fits our purposes. The freedom in having different nonflat Hilbert--space bundles 
over ${\bf CP}^n$ resides in the different possible choices for the complex line bundle 
$N({\bf CP}^n)$. Such choices are 1--to--1 with the elements of the Picard group 
${\rm Pic}\,({\bf CP}^n)={\bf Z}$. The latter appears as the parameter space 
for physically inequivalent choices of the vacuum state. Every choice of a 
vacuum leads to a different set of excitations and thus to a different 
quantum mechanics. Moreover, the ${\cal QH}$--bundles constructed here are nonflat. 
This implies that, even after fixing a vacuum,
there is still room for duality transformations between different observers 
on classical phase space. These two facts provide an explicit 
implementation of quantum--mechanical dualities.

{\bf Acknowledgements}

It is a great pleasure to thank Prof. J. A. de Azc\'arraga for encouragement 
and support, and the organisers  of the ``Azc\'arragafest" for the invitation 
to participate. This work has been partially supported by research grant BFM2002--03681 
from Ministerio de Ciencia y Tecnolog\'{\i}a and EU FEDER funds.

\end{document}